\documentstyle[12pt,aasms4]{article}
\begin{document}
\title{A Survey for Low-Surface-Brightness Galaxies Around
M31.\ I.\ The Newly Discovered Dwarf Andromeda V}
\author{Taft E.\ Armandroff}
\affil{Kitt Peak National Observatory, National Optical Astronomy
Observatories\altaffilmark{1},\\ P.O.\ Box 26732, Tucson, AZ 85726}
\authoremail{armand@noao.edu}
\author{James E.\ Davies}
\affil{Kitt Peak National Observatory, National Optical Astronomy
Observatories\altaffilmark{1},\\ P.O.\ Box 26732, Tucson, AZ 85726;\\
and\\ Department of Astronomy, University of Wisconsin, Madison,
WI 53706}
\authoremail{davies@uwast.astro.wisc.edu}
\and
\author{George H.\ Jacoby\altaffilmark{2}}
\affil{Kitt Peak National Observatory, National Optical Astronomy
Observatories\altaffilmark{1},\\ P.O.\ Box 26732, Tucson, AZ 85726}
\authoremail{jacoby@noao.edu}
\altaffiltext{1}{The National Optical Astronomy Observatories are
operated by AURA, Inc., under cooperative agreement with the National
Science Foundation.}
\altaffiltext{2}{Visiting Astronomer, Steward Observatory, University
of Arizona}
\slugcomment{Accepted by The Astronomical Journal, November 1998 issue}
\lefthead{Armandroff, Davies \& Jacoby}
\righthead{Low-Surface-Brightness Galaxies Around M31}

\begin{abstract}
We present images and a color--magnitude diagram for And V, a new dwarf
spheroidal companion to M31 that was found using a digital filtering
technique applied to 1550 deg$^2$ of the second Palomar Sky Survey.
And V resolves into stars easily in follow-up 4-m $V$- and $I$-band
images, from which we deduce a distance of 810 $\pm$ 45 kpc using the
tip of the red giant branch method. Within the uncertainties, this
distance is identical to the Population II distances for M31 and,
combined with a projected separation of 112 kpc,  provides strong
support for a physical association between the two galaxies. There is
no emission from And V detected in H$\alpha$, 1.4 GHz radio continuum,
or IRAS bandpasses, and there is no young population seen in the
color--magnitude diagram that might suggest that And V is an
irregular.  Thus, the classification as a new dwarf spheroidal member
of the Local Group seems secure.  With an extinction-corrected central
surface brightness of 25.2 $V$ mag/arcsec$^2$, a mean metal abundance
of [Fe/H] $\approx$ --1.5, and no evidence for upper AGB stars, And V
resembles And I \& III.
\end{abstract}

\section{Introduction}
A complete and accurate knowledge of dwarf galaxies in the Local Group
is necessary for many problems in galaxy formation and cosmology (see
Ferguson \& Binggeli 1994, Gallagher \& Wyse 1994).  For example, the
faint end of the luminosity function in large clusters of galaxies is
more populated than that of the Local Group (Trentham 1998).  Because
this result may be explained by incompleteness and/or small number
statistics in the Local Group, it remains unknown whether
low-luminosity galaxies form more frequently in denser environments or
not.  As another example, van den Bergh (1994) found a correlation
between the Galactic dwarfs' galactocentric distances and the
prominence of an intermediate-age population. This may reflect the
effects of ram pressure stripping or photoionization via the early
Milky Way's integrated UV flux on the inner dwarf spheroidals.
However, our ability to confirm or rule out this trend in the
independent case of M31 is hampered by its small number of known
companions.

The Milky Way has nine low-surface-brightness gas-poor companions;
these are usually called dwarf spheroidal galaxies. Dwarf spheroidals
are now considered low-luminosity dwarf elliptical galaxies.  The
empirical correlations followed by dwarf spheroidals join smoothly onto
the dwarf elliptical correlations (e.g., central surface brightness
vs.\ absolute magnitude, mean metal abundance vs.\ absolute
magnitude).  M31 has three dwarf spheroidal companions (And I, II \&
III) and three dwarf elliptical companions (NGC 147, 185 \&
205)\footnote{See Gallagher \& Wyse (1994), Armandroff (1994), and Da
Costa (1998) for recent reviews on the dwarf companions to the Milky
Way and M31.}.  This relatively small number would seem to suggest a
deficit of dwarf spheroidal companions of M31 compared to the Milky
Way.  Is this a real effect or simply due to the fact that the Milky
Way has been searched more thoroughly than M31?  Alternatively, it is
possible that M31's environment was more conducive to producing more
luminous dwarfs like NGC 147, 185 and 205, rather than very low
luminosity dwarfs.  Another possibility is that M31 was more effective
than the Milky Way in disrupting and accreting its original population of
low-luminosity dwarfs.

The current census of M31 dwarf spheroidals resulted from the
pioneering survey of van den Bergh (1972a, 1974).  He visually searched
$\sim$700 square degrees around M31 using IIIaJ photographic plates
taken with the Palomar 48-inch Schmidt telescope.  Three dwarf
spheroidal candidates were found (And I, II \& III).  One smaller,
higher surface brightness object was also found (And IV); subsequent
work confirmed van den Bergh's (1972a) conclusion that it is not a
dwarf spheroidal (Jones 1993).  And I, II \& III were found to resolve
into stars at approximately the same magnitude as the M31 companion NGC
185 (van den Bergh 1972b, 1974), supporting the notion that these three
galaxies are physically associated with M31.  As more precise distances
to And I, II \& III have become available (e.g., Armandroff et
al.\ 1993, Da Costa et al.\ 1996), the conclusion that these galaxies
are located in the outer reaches of M31's halo has remained valid.

For the Galactic dwarf spheroidals, all three components of their
galactocentric distance are known.  However, only projected distances
from the center of M31 are known for And II \& III (see Da Costa et
al.\ 1996 for a precise line-of-sight distance for And I, based on the
magnitude of the horizontal branch, and the resulting limits on its
true three-dimensional distance from the center of M31).  Figure 6 of
Armandroff (1994) compares the projected galactocentric distance
distribution of the M31 dwarf spheroidals with that of their Galactic
counterparts.  Van den Bergh's (1972a, 1974) survey begins to lack
complete areal coverage at a projected galactocentric distance of
$\sim$110 kpc, and it has no coverage beyond a projected galactocentric
distance of $\sim$260 kpc.  Since Leo I and II are located $>$ 200 kpc
from the Galactic center, some M31 dwarfs with very large
galactocentric distances may be outside van den Bergh's survey area.
Finally, because And III with a central surface brightness of 25.3 $V$
mag/arcsec$^2$ was found by van den Bergh, we know that his survey was
sensitive to quite low surface brightness.  However, the central
surface brightness to which van den Bergh's (1972a, 1974) survey is
complete is not known.

In order to find possible additional M31 low-surface-brightness
companions, we undertook a new search of the area around M31.  Our
search uses the Second Palomar Sky Survey (POSS-II; Reid et al.\ 1991,
Reid \& Djorgovski 1993), which has higher resolution and extends
substantially deeper than its predecessor.  The availability of the
POSS-II in digital form (Lasker \& Postman 1993) allows the use of
digital image processing techniques.  These techniques may enable the
detection of dwarf galaxies of even lower central surface brightness than
van den Bergh's (1972a, 1974) completeness limit.  The digital POSS-II
also facilitates the search of an arbitrarily large area of sky around M31.

This paper will provide a brief description of our search strategy and
techniques.  This will be followed by the first results of our survey,
in which we identify a new dwarf galaxy associated with M31.

\section{Search Methodology}

The POSS-II data used for the survey consists of red IIIaF $+$ RG610
plates that were digitized at STScI.  The red bandpass was chosen for
searching for dSph galaxies because most of their light is typically
emitted by red giants and red horizontal branch stars.  The fainter sky
and broader bandpass of the blue plates suggest that these might have
similar detection efficiency, but at the time of our study only the red
plates were accessible digitally.  The plates are 6.6 degrees on a side
with 6.3 degrees of useable emulsion.  They are spaced 5 degrees
between plate centers which gives a 1.3 degree overlap area for
cross-checking candidates to eliminate spurious detections.  Having
this relatively large overlap is useful considering the optical
vignetting effects in the outer regions of the plates as discussed by
Reid et al.\ (1991).  Each digitized plate is subdivided into a grid of
7$\times$7 1 deg$^2$ images (each $\sim$3564 pixels on a side) to make
the image processing more manageable.  The 49 images per plate
consequently have overlaps of 7 arcmin which is larger than the FWHM of
a typical dwarf galaxy profile at the distance of M31; therefore no
objects are missed due to being close to an edge.

Each 1 deg$^2$ image is filtered separately in a four-step process for
the purpose of increasing the signal-to-noise of low-surface-brightness
(LSB) features.  The procedure was optimized for the detection of And
II \& III.  These steps are:

\begin{enumerate}

\item A fourth order polynomial is fit to and subtracted from the
background of each image in order to make the background as
flat as possible.  It should be noted that these plates are remarkably
flat on scales of several cm, corresponding to about a degree.  The
background needs to be as flat as possible due to the way that
high-surface-brightness (HSB) objects are removed in step 2.  It was found
through successive trials of fitting polynomials of different order
that a fourth order polynomial flattened the plate well enough without
running the risk of removing LSB features on the scale of a few
arcmin.  Values 2$\sigma$ higher than the image median are not used in
the fitting so that the polynomial fit is dominated by the background
and not by stars and galaxies in the image (where $\sigma$ is the
standard deviation).

\item Stars on the images are removed by a clipping operation that
calculates the median value of the full image and replaces all values
$> 0.75\sigma$ above and $< 0.75\sigma$ below the median with the
median value.  The positive clip has the effect of removing all
HSB objects---stars, obvious galaxies and clusters of
galaxies---that hamper the detection process.  The negative clip gets
rid of darkened rings caused by dust and other plate defects.  The
clipping operation works well since it has no preference for the size
or shape of the HSB object---it removes everything.  The value of
$0.75\sigma$ was found to be optimal in that it removes much of the
``wings'' of large, bright stars but does not clip too low and run the
risk of diminishing the peaks of our test dSph galaxies, And II \&
III.

\item The images are then filtered with a square spatial median filter
of size 77$\times$77 arcsec that moves across the image replacing the
central pixel value with the median of the 77$\times$77 pixel values
contained in the box.  Davies et al.\ (1994) found that in order to
optimally enhance and detect an LSB disk galaxy, their
cross-correlation filter size and the scale length of the galaxy ought
to be roughly equal within a factor of 2.  We confirm this for a median
filter and LSB dwarf galaxies and find that for dSph with exponential
scale lengths in the 45 to 94 arcsec range defined by the known
Andromeda dwarfs (Caldwell et al.\ 1992), an optimal median filter size
is 77$\times$77 arcsec.

\item Detection of LSB dwarf galaxies is carried out by visually
comparing the filtered images with the raw images and selecting
candidates that resemble And II \& III on the raw and processed
images.  LSB features are prominent on the processed images but are
extremely weak, and in many cases invisible, on the unprocessed
images.  And II \& III are trivial to detect on processed POSS-II
images (processed POSS-I images easily reveal all three M31 dSph).
Images of And II \& And III from the POSS-II, raw and after processing,
are shown in Figure \ref{and2and3dss} (And I is not shown because the
relevant digitized POSS-II image is not yet available).
\placefigure{and2and3dss}
\end{enumerate}

The survey area was selected based on the area searched by van den
Bergh (1974) and was extended out 6--10 degrees farther in all directions
for the completeness reasons stated in the Introduction.  Figure
\ref{survey_map} shows the area that has been searched so far as shaded
boxes, with the van den Bergh area outlined.  The area searched
consists of 51 plates that cover approximately 1550 deg$^2$.
\placefigure{survey_map}

Since the image processing enhances any density feature on the plate
that is the proper size and low enough surface brightness (a few
percent above sky), and because the survey area is very large,
identifying and eliminating contaminants that masquerade as LSB dwarfs
is a daunting task.  The biggest offenders are ghosts produced from the
reflection of bright stars in the telescope optics and Galactic
reflection nebulae (also referred to as Galactic cirrus clouds; see
Bremmes, Binggeli \& Prugniel 1998).  In many cases follow-up
observations are the only way to eliminate candidates.  Fortunately,
many detections are asymmetric and lack a smooth exponential profile,
rendering them inconsistent with an LSB dwarf galaxy signature.
Therefore, such objects can be eliminated from the candidate list.  The
remainder, however, require follow-up CCD imaging, which is described
in the next section.

\section{Follow-Up Observations}

The remainder of this paper will concentrate on one excellent candidate
that was found at the following celestial coordinates: $\alpha_{2000}$
= 1:10:17.1, $\delta_{2000}$ = +47:37:41.  These correspond to Galactic
coordinates of: $l$ = 126.2$^\circ$, $b$ = --15.1$^\circ$.  There are
no previously catalogued galaxies/objects at these coordinates
according to the NASA/IPAC Extragalactic Database\footnote{The
NASA/IPAC Extragalactic Database (NED) is operated by the Jet
Propulsion Laboratory, California Institute of Technology, under
contract with the National Aeronautics and Space Administration.} and
the SIMBAD database.  Because other ``newly found" nearby dwarf
galaxies have occasionally turned out to be rediscoveries, we have
inspected a number of lists that may contain such galaxies in order to
verify that this object is indeed new, including Karachentseva \&
Karachentsev (1998), Nilson (1973), Schmidt \& Boller (1992), Ellis et
al.\ (1984), Weinberger (1995), and Schombert et al.\ (1992).
Following van den Bergh (1972a), we will call this new system Andromeda
V.  Figure \ref{survey_map} shows where And V falls in the search
area.

Figure \ref{and5dss} shows And V on the digitized POSS-II, raw and
after processing.  One sees a strong resemblance between And V and the
known M31 dwarf spheroidals And II \& III on the raw and processed
POSS-II images (see Figure \ref{and2and3dss}).  We examined the POSS-I
prints and found that And V is not visible on them; And I, II \& III
are not clearly visible on the POSS-I prints either.  And I \& II are
easily visible on the POSS-II transparencies.  And III, which is of
lower central surface brightness than And I \& II (Caldwell et
al.\ 1992), is just visible on the POSS-II.  And V appears to be of
lower surface brightness still and is marginally discernible on the
POSS-II provided that the coordinates are known a priori.
Karachentseva \& Karachentsev (1998) also noted that And I, II \& III
are visible on the POSS-II.
\placefigure{and5dss}

The first step in clarifying the nature of a candidate from the sky
survey is small-telescope CCD imaging.  Such imaging eliminates
candidates that are really ghosts from bright stars  on the POSS-II.
Other potential ``contaminants" that can be eliminated via
small-telescope imaging include: distant clusters of galaxies,
planetary nebulae, and distant low-surface-brightness
spirals.\footnote{Note that we have not completed follow-up imaging of
all of our candidate LSB dwarfs.  We will report follow-up imaging of
other candidates in a future paper.}  And V was imaged at the KPNO
0.9-m telescope with the CCD Mosaic Imager (see Armandroff et
al.\ 1998) in the $R$ band for 600 seconds on September 28, 1997.  The
scale was 0.43 arcsec/pixel, and the seeing was 1.4 arcsec FWHM.  In
the 0.9-m data, And V looks like a low-surface-brightness galaxy and
shows incipient resolution into stars.

And V was imaged at the KPNO 4-m telescope prime focus with the
T2KB CCD on November 2, 1997 (see Massey et al.\ 1997 for a description
of the imager).  The scale was 0.42 arcsec/pixel, and the seeing was
1.0 arcsec FWHM.  Three $V$ exposures of 900 sec each and seven $I$
exposures of 300 sec each were obtained.  It was photometric and $\sim$15
Landolt (1992) standards in three fields were observed.  The combined
4-m $V$ image is displayed in Figure \ref{and5ccd} (the central 1000
$\times$ 1000 pixels).  And V resolves nicely into stars on both the
$V$ and the $I$ images.  One sees a smooth stellar distribution.  In
these images, And V resembles the other M31 dwarf spheroidals (see
Figure 1 of Mould \& Kristian 1990 and/or Figure 1 of Armandroff et
al.\ 1993).  And V does not look lumpy or show obvious regions of star
formation that would suggest a dwarf irregular classification.  No
obvious globular clusters are seen in And V, though none are expected
for a low-luminosity dwarf spheroidal (e.g., Sec.\ 4.3 of Da Costa \&
Armandroff 1995).
\placefigure{and5ccd}

In order to look for possible ionized gas, And V was observed with the
KPNO 4-m telescope in H$\alpha$ narrow-band and $R$ on October 21, 1997
with the CCD Mosaic Imager.  The H$\alpha$ filter used has a central
wavelength of 6569 \AA\ and a width of 80 \AA\ FWHM in the f/3.2 beam
of the telescope.  Five narrow-band exposures of 600 sec each and five
$R$ exposures of 400 sec each were obtained.  The average seeing was
1.2 arcsec FWHM, sampled at 0.26 arcsec/pixel.  We computed scaling
ratios between the combined H$\alpha$ and $R$ images and performed the
subtraction. Figure \ref{and5halpha} shows both the H$\alpha$ image and
the continuum-subtracted H$\alpha$ image.  No diffuse H$\alpha$
emission or H {\sc ii} regions were detected in And V.  Some residuals
are present in this image due to the somewhat different behavior of the
PSF in $R$ and H$\alpha$.  There are a few small positive residuals,
but these are all detected strongly in $V$, $R$, and $I$ and therefore
are not H {\sc ii} regions or planetary nebulae.  Their slightly
extended nature suggests that they are distant galaxies.  The lack of
H$\alpha$ emission in And V likely rules out significant current star
formation, reinforcing our conclusion, based on And V's appearance on
the broad-band images, that it is a dwarf spheroidal galaxy rather than
a dwarf irregular.  Almost all Local Group dwarf irregulars are easily
detected in H$\alpha$ (e.g., Kennicutt 1994).
\placefigure{and5halpha}

We also examined the IRAS maps of the And V region, using the FRESCO
data product from IPAC\footnote{IPAC is funded by NASA as part of the
IRAS extended mission under contract to JPL.}.  And V is not detected
in any of the IRAS far-infrared bands.  And I, II \& III are not
detected by IRAS either.  Far-infrared emission, as seen by IRAS, is
the signature of warm dust.  Some Local Group dwarf irregular galaxies,
such as NGC 6822, IC 1613 and WLM, are detected by IRAS (Rice et
al.\ 1988).  Hence the non-detection of And V by IRAS is further
evidence that it does not contain substantial amounts of dust and gas.
However, many of the less active, less luminous Local Group dwarf
irregulars are not detected by IRAS either.  Consequently, the lack of
far-infrared emission from And V is probably a weaker constraint on the
presence of an interstellar medium than the lack of H$\alpha$
emission.

And V is not detected in the NRAO VLA 1.4 GHz radio continuum survey
(Condon et al.\ 1998).  This is as expected because none of the Local
Group dwarf spheroidals are detected in this survey, and even most
Local Group dwarf irregulars are not detected.  IC 10 is easily
detected in the 1.4 GHz survey, but it has an exceptional
star-formation rate and disturbed interstellar medium (Yang \& Skillman
1993, Massey \& Armandroff 1995).

We know nothing about the H {\sc i} content of And V via 21 cm
observations.  Either an H {\sc i} detection or a strict upper limit
would be valuable.

\section{Color--Magnitude Diagram}

The next step in ascertaining the nature of And V is the construction
of a color--magnitude diagram for its member stars  in order to reveal
its distance and stellar populations characteristics.
Instrumental magnitudes were measured on the And V 4-m images using the
IRAF implementation of the {\sc daophot} crowded-field photometry program
(Stetson 1987, Stetson et al.\ 1990).  The standard {\sc daophot} procedure
was used, culminating in multiple iterations of {\sc allstar}.  Because
substantial point-spread-function variations occur between the middle
and corners of the CCD frames, only the central $1000 \times 1000$
region was analyzed.  Stars with anomalously large values of the
{\sc daophot} parameter CHI were removed from the photometry lists.

Large-aperture photometry was performed for the $\sim$15 well-exposed
Landolt (1992) photometric standard stars.  Assuming mean atmospheric
extinction coefficients for Kitt Peak, photometric transformation
equations were derived.  Besides the zeropoint, only a linear color
term of small size was needed in each filter.  The And V {\sc allstar}
magnitudes were placed on the standard system via large-aperture
photometry of several bright, uncrowded stars on the And V images in
each filter.

Initial color--magnitude diagrams for the parts of the images dominated
by And V stars revealed a red giant branch, which is absent in the
outer regions of the images.  We used the spatial distribution of the
stars in the red-giant region of the color--magnitude diagram to
determine a reasonably precise center for And V (see Figure
\ref{and5ccd})\footnote{The x,y coordinates of this center were
transformed to celestial coordinates using the HST Guide Star Catalog.
This is the origin of the celestial coordinates for And V given in the
first paragraph of Sec.\ 3.  The uncertainty in the coordinates is
dominated by the error in determining the centroid of likely member
giants and is roughly 5 arcsec.}.

The color--magnitude diagram of And V is illustrated in Figure
\ref{2panelcmd}.  The left panel shows the color--magnitude diagram for
the region where And V stars dominate over field contamination, a
circle of radius 170 pixels or 71 arcsec centered on And V (chosen as a
compromise between maximizing the number of And V stars and minimizing
field contamination; hereafter referred to as the And V region).  The
upper red giant branch of And V is visible in this left panel.  The
right panel displays the color--magnitude diagram for the portion of
the frames where the And V contribution is negligible, that area
outside a radius of 400 pixels or 168 arcsec from the center of And V
(hereafter referred to as the field region).  The And V giant branch is
absent from the right panel.  Instead, a broad swath of field stars
covers most of the color--magnitude plane.  Considering that the field
region represents 5.5 times more area than the And V region, most of
the stars that deviate from red giant branch in the left panel can be
interpreted as field contamination.  As a rough comparison, the colors
and magnitudes of the And V stars in the left panel of Figure
\ref{2panelcmd} are similar to those of the And III bright red giants
(see Figure 5 of Armandroff et al.\ 1993).  This similarity, which we
will investigate carefully below, substantially strengthens the
evidence that And V is indeed associated with M31.
\placefigure{2panelcmd}

A distance can be derived for And V based on the $I$ magnitude of the
tip of the red giant branch (Da Costa \& Armandroff 1990, Lee et
al.\ 1993).  Figure \ref{and5_lf} shows $I$ luminosity functions for
the And V region and for the field region (the same regions as used in
constructing the color--magnitude diagrams in Figure \ref{2panelcmd}).
Based on the magnitude at which the And V luminosity function begins to
rise strongly above that of the field, the $I$ magnitude of the red
giant branch tip is 20.85 $\pm$ 0.10; this value is supported by the
apparent location of the tip in the color--magnitude diagram (Figures
\ref{2panelcmd} and \ref{cmd_fiducial}).  For metal-poor systems such as
And V (see below), the red giant branch tip occurs at $M_I$ = --4.0 (Da
Costa \& Armandroff 1990, Lee et al.\ 1993).  We adopt an interstellar
reddening for And V of E($B$--$V$) = 0.16 $\pm$ 0.03 from the Burstein
\& Heiles (1982) extinction maps.  Assuming $A_V$ =
3.2$\times$E($B$--$V$) and the E($B$--$V$) to E($V$--$I$) conversion of
Dean, Warren \& Cousins (1978), E($V$--$I$) = 0.21 and $A_I$ = 0.30.
The above values then yield a true distance modulus for And V of 24.55
$\pm$ 0.12 and a distance of 810 $\pm$ 45 kpc.  These error estimates
do not include any systematic uncertainty in the zeropoint of the
adopted distance scale.
\placefigure{and5_lf}

How does this And V distance compare with that of M31?  The most
directly comparable distance determinations for M31 are based on either
red giant branch tip stars or RR Lyraes in the M31 halo or horizontal
branch stars in M31 globular clusters.  Such distance determinations
are discussed in Da Costa et al.\ (1996) in order to compare the
distances of M31 and And I.  On the same distance scale as used here,
the red giant branch tip stars (Mould \& Kristian 1986) and RR Lyraes
(Pritchet \& van den Bergh 1988) agree well and yield an M31 distance
of 760 $\pm$ 45 kpc.  On the other hand, the horizontal branch
magnitudes for eight globular clusters observed by HST (Fusi Pecci et
al.\ 1996) suggest an M31 distance of 850 $\pm$ 20 kpc on the same
scale.  The Cepheid distance to M31 is 770 kpc (Freedman \& Madore
1990); Lee et al.\ (1993) and Sakai et al.\ (1997) have shown that the
Cepheid distance scale and the tip of the red giant branch scale are
consistent.  Our distance for And V of 810 $\pm$ 45 kpc implies that
And V is located at the same distance along the line of sight as M31 to
within the uncertainties.  And V's projected distance from the center
of M31 is 112 kpc; And I, II \& III have projected M31-centric
distances of 46, 144 and 69 kpc, respectively.  The above line-of-sight
and projected distances strongly suggest that And V is indeed
associated with M31.

In order to investigate the stellar populations in And V, we have
overplotted its color--magnitude diagram with fiducials representing
the red giant branches of Galactic globular clusters that span a range
of metal abundance (Da Costa \& Armandroff 1990).  For an old stellar
population (as demonstrated below for And V), the color of the red
giant branch is primarily determined by metal abundance.  Figure
\ref{cmd_fiducial} shows $I$,$V$--$I$ color--magnitude diagrams for the
And V  and field regions, with fiducials for M15 ([Fe/H] = --2.17), M2
(--1.58), NGC 1851 (--1.16), and 47 Tuc (--0.71, dotted line).  Most of
the And V giants lie near or slightly redwards of the M2 fiducial.
Thus, the mean metal abundance of And V is approximately --1.5.  This
metallicity is normal for a dwarf spheroidal (e.g., see Figure 9 of
Armandroff et al.\ 1993).
\placefigure{cmd_fiducial}

In Figure \ref{cmd_fiducial}, the region of the color--magnitude
diagram substantially blueward of the lowest metallicity fiducial (M15)
is empty.  Since the $V$ data reach significantly deeper than the $I$
data, stars in this region of the color--magnitude diagram should be
easily detected.  Thus, no bright blue stars appear to be present in
And V.  Using the Bertelli et al.\ (1994) isochrones, we conservatively
estimate that no stars younger than at least 200 Myr are present in the
And V region of Figure \ref{cmd_fiducial}.  Local Group dwarf irregular
galaxies contain many stars younger than 200 Myr.  As a conservative
example, LGS 3, which is considered a transition object with both dwarf
spheroidal and dwarf irregular characteristics, has stars of age 100
Myr (Mould 1997).  Thus, the lack of young stars is strong evidence
that And V is a dwarf spheroidal and not a dwarf irregular.  We hope to
obtain deeper imaging of And V, particularly in the $B$ band, in order
to place stronger limits on the youngest stars in And V.

Using the luminosities and numbers of upper asymptotic giant branch
(AGB) stars in a metal-poor stellar system, one can infer the age and
strength of its intermediate-age component (Renzini \& Buzzoni 1986).
Armandroff et al.\ (1993) have analyzed the luminosity functions of And
I \& III and deduced that the fraction of each galaxy's luminosity
contributed by an intermediate-age population is 10 $\pm$ 10 percent
(where intermediate age denotes 3--10 Gyr old on an age scale where the
Galactic globular clusters are $\approx$14 Gyr old).  Because And II
contains upper AGB carbon stars (Aaronson et al.\ 1985), its
intermediate-age component seems more substantial (though a
quantitative analysis has not been performed).  Intermediate-age
components of varying prominence are present in the Galactic dwarf
spheroidals, ranging from $\sim$75\% of all stars in Carina to
$\sim$0\% in Ursa Minor (e.g., Da Costa 1998).  Armandroff et
al.\ (1993) used upper AGB stars with $M_{\rm bol}$ values between
--3.8 and --4.6 to constrain the number of stars with ages between 3
and 10 Gyr.  Observations of And I revealed 4 $\pm$ 4 such stars; And
III contains 3 $\pm$ 2.9 such stars.  For And V, the above $M_{\rm
bol}$ limits correspond to 19.75 $\leq$ $I$ $<$ 20.45.  Examination of
the And V luminosity function (Figure \ref{and5_lf}) indicates no
significant excess of And V stars over the field in this magnitude
interval.  After appropriate color limits and field subtraction, there
are 0.9 $\pm$ 3.0 excess stars in the And V region within the upper AGB
magnitude limits.  Thus, And V does not contain significant numbers of
upper AGB stars.  Therefore, it does not have a prominent
intermediate-age population; in this sense, it is similar to And I \&
III.

Because dwarf spheroidal galaxies are faint compared to the terrestrial
night sky, investigation of their surface brightness profiles requires
very careful attention to flat fielding and background subtraction (see
Sec.\ 3.1 of Caldwell et al.\ 1992).  The current data are not well
suited to determination of And V's surface brightness profile.
However, it is possible to derive And V's central surface brightness
via large-aperture photometry in the core of And V.  As a first step,
we subtracted all stars brighter than $V$ = 22 from the And V $V$
image, since they are field contamination.  We then performed digital
aperture photometry with an aperture radius of 20 arcsec and a sky
annulus near the edge of the image (193--204 arcsec).  This yielded an
And V apparent $V$ central surface brightness of 25.7 mag/arcsec$^2$.
This result is insensitive to the precise specification of the sky
annulus, and changing the radius of the aperture by 5 arcsec results in
surface brightness changes at the 0.1 mag level.  And V has a fainter
apparent central surface brightness than And I, II \& III (24.9, 24.8,
and 25.3 $V$ mag/arcsec$^2$, respectively; Caldwell et al.\ 1992).  And
V probably eluded detection until now due to its very low apparent
surface brightness.  Our And V measurement corresponds to an
extinction-free $V$ central surface brightness of 25.2 mag/arcsec$^2$.

\section{Discussion}
The discovery of And V increases the number of M31 dwarf spheroidals
from three to four.  It changes somewhat the properties of M31's
satellite system, as discussed below.  A more complete description of
M31's entourage of dwarf spheroidals will be presented once our survey
and the necessary follow-up observations are complete.

The spatial distribution of the companions to M31 has been discussed by
Karachentsev (1996).  The discovery of And V changes somewhat the
spatial distribution of the M31 satellites.  The location of And V
relative to M31, M32, M33, NGC 147, 185 \& 205, and And I, II \& III is
shown in Figure \ref{survey_map}.  Curiously, And I, II \& III are all
located south of M31, while the three more luminous dwarf elliptical
companions NGC 147, 185 \& 205 are all positioned north of M31.  Also,
Karachentsev (1996) notes that there are more M31 companions overall
south of M31 than north of M31.  And V's location north of M31 lessens
both of these asymmetries.  Karachentsev (1996) called attention to the
elongated shape of the M31 companion spatial distribution (based on the
previously known galaxies labelled in Figure \ref{survey_map}, plus the
more distant galaxies IC 1613, IC 10, and LGS 3, which he argues are
also associated with M31) with axis ratio 5:2:1.  And V is located
within this flattened spatial distribution.  With a projected radius
from the center of M31 of 112 kpc, And V increases slightly the mean
projected radius of the M31 dwarf spheroidals from 86 kpc to 93 kpc.
Finally, Karachentsev (1996) also discussed morphological
segregation among the M31 companions, with the dwarf ellipticals and
dwarf spheroidals located closer to M31, and the spirals and
irregulars on the outskirts (see also van den Bergh 1994).  And V's
distance from M31 and our classification of this galaxy as a dwarf
spheroidal support the presence of morphological segregation.

The discovery of nearby dwarf galaxies like And V augments the faint
end of the luminosity function of the Local Group.  We do not yet have
a reliable $M_V$ value for And V, but it appears to be similar to that
of And III ($M_V$ = --10.2) since they have similar
extinction-corrected central surface brightness.  From a survey of nine
clusters of galaxies, Trentham (1998) derived a composite luminosity
function that is steeper at the faint end than that of the Local Group
(see his Figure 2).  He attributed the difference to poor counting
statistics and/or incompleteness among the Local Group sample.  The
discovery of And V reduces slightly the discrepancy between the Local
Group luminosity function and the extrapolation of Trentham's (1998)
function.  However, many more galaxies would be required to equate the
two functions (for example, the difference is 29 $\pm$ 14 galaxies in
Trentham's faintest two bins, --13 $\leq$ $M_B$ $\leq$ --11).

Based on And V's smooth stellar distribution, lack of H$\alpha$ and
IRAS emission, and absence of young stars in the color--magnitude
diagram, we conclude that And V is a dwarf spheroidal galaxy.  A deeper
color--magnitude diagram, featuring a bluer color like $B$, and an H
{\sc i} search would allow us to evaluate more carefully whether And V
could be a transition object between dwarf spheroidals and dwarf
irregulars, like LGS 3 but older (see Lee 1995, Mould 1997, Young \& Lo
1997).  Based on the available data however, And V is dwarf spheroidal
companion to M31, and its stellar population is not obviously different
from those of And I or III.

\acknowledgments
The Digitized Sky Surveys were produced at the Space Telescope Science
Institute under U.S.\ Government grant NAG W-2166.  The survey images
used here are based on photographic data obtained using the Oschin
Schmidt Telescope on Palomar Mountain.  The Second Palomar Observatory
Sky Survey was made by the California Institute of Technology with
funds from the National Science Foundation, the National Geographic
Society, the Sloan Foundation, the Samuel Oschin Foundation, and the
Eastman Kodak Corporation.  The Oschin Schmidt Telescope is operated by
the California Institute of Technology and Palomar Observatory.
Supplemental funding for sky-survey work at the STScI is provided by
the European Southern Observatory.  We appreciate the efforts of Barry
Lasker and Jesse Doggett at STScI and Watanabe Masaru at the National
Astronomical Observatory of Japan in helping us to access the DSS
data.  We are grateful to Nelson Caldwell, Gary Da Costa, and Jay Gallagher
for comments on an earlier version of this manuscript.  JED was supported
by the NSF Research Experiences for Undergraduates Program at NOAO
during the summer of 1997.  GHJ wishes to thank Peter Strittmatter for
providing a sabbatical office at Steward Observatory during the period
when this paper was written.

\begin{figure}[p]
\caption[]{The left panels show images of And II \& III from the
digitized POSS-II.  The right panels show the results of applying our
digital enhancement procedure to these POSS-II images.  The signatures
of the galaxies are much more clearly visible in the processed images.
The And III processed image also shows a ``doughnut" artifact resulting
from a very bright, saturated star.  Each panel is 17 arcmin on a
side.  North is at the top, and East is to the left.}
\label{and2and3dss}
\end{figure}

\begin{figure}[p]
\plotone{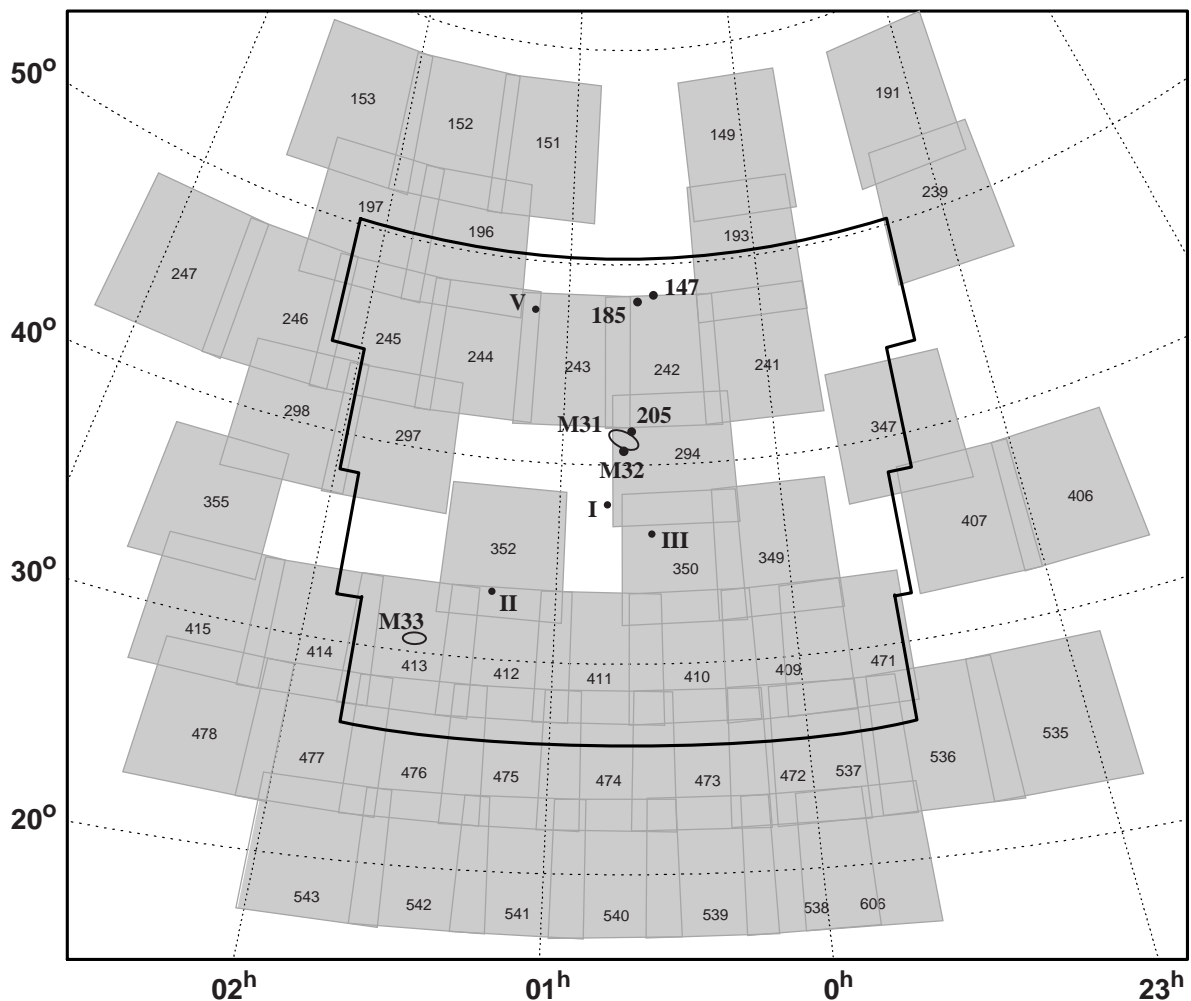}
\caption[]{Map of the region of sky around M31 that has been surveyed.
Individual POSS-II plates that have been searched are shown and
labelled.  The area surveyed by van den Bergh (1972a, 1974) is
thick-outlined.  M31 and its known neighbors are labelled.  The new M31
dwarf spheroidal, And V, is also indicated.}
\label{survey_map}
\end{figure}

\begin{figure}[p]
\caption[]{The left panel shows an  image of And V from the digitized
POSS-II.  The right panel shows the effect of applying our digital
enhancement procedure to the POSS-II image. Each panel is 17 arcmin on
a side.  North is at the top, and East is to the left.  Note the
resemblance between And V and the Figure \ref{and2and3dss} images of
And II \& III.}
\label{and5dss}
\end{figure}

\begin{figure}[p]
\caption[]{Image of And V made from the combination of three 900-second
exposures through the $V$ filter with the KPNO 4-m telescope.  North is
at the top, and East is to the right.  The image is 7 arcmin (1000
pixels) on a side; And V's center is at $x \approx 500$, $y \approx 514$.}
\label{and5ccd}
\end{figure}

\begin{figure}[p]
\caption[]{An H$\alpha$ image of And V made from the combination of
five 600-second exposures with the KPNO 4-m telescope is shown in the
left panel.  A continuum-subtracted version of this image is displayed
in the right panel.  In each panel, North is at the top, and East is to
the right.  Each image is 2.7 arcmin on a side.  The residuals in the
subtracted image around many of the brighter stars are due to small
differences in the PSF between H$\alpha$ and $R$.  Note the lack of
diffuse H$\alpha$ emission or H {\sc ii} regions.}
\label{and5halpha}
\end{figure}

\begin{figure}[p]
\epsscale{0.85}
\plotone{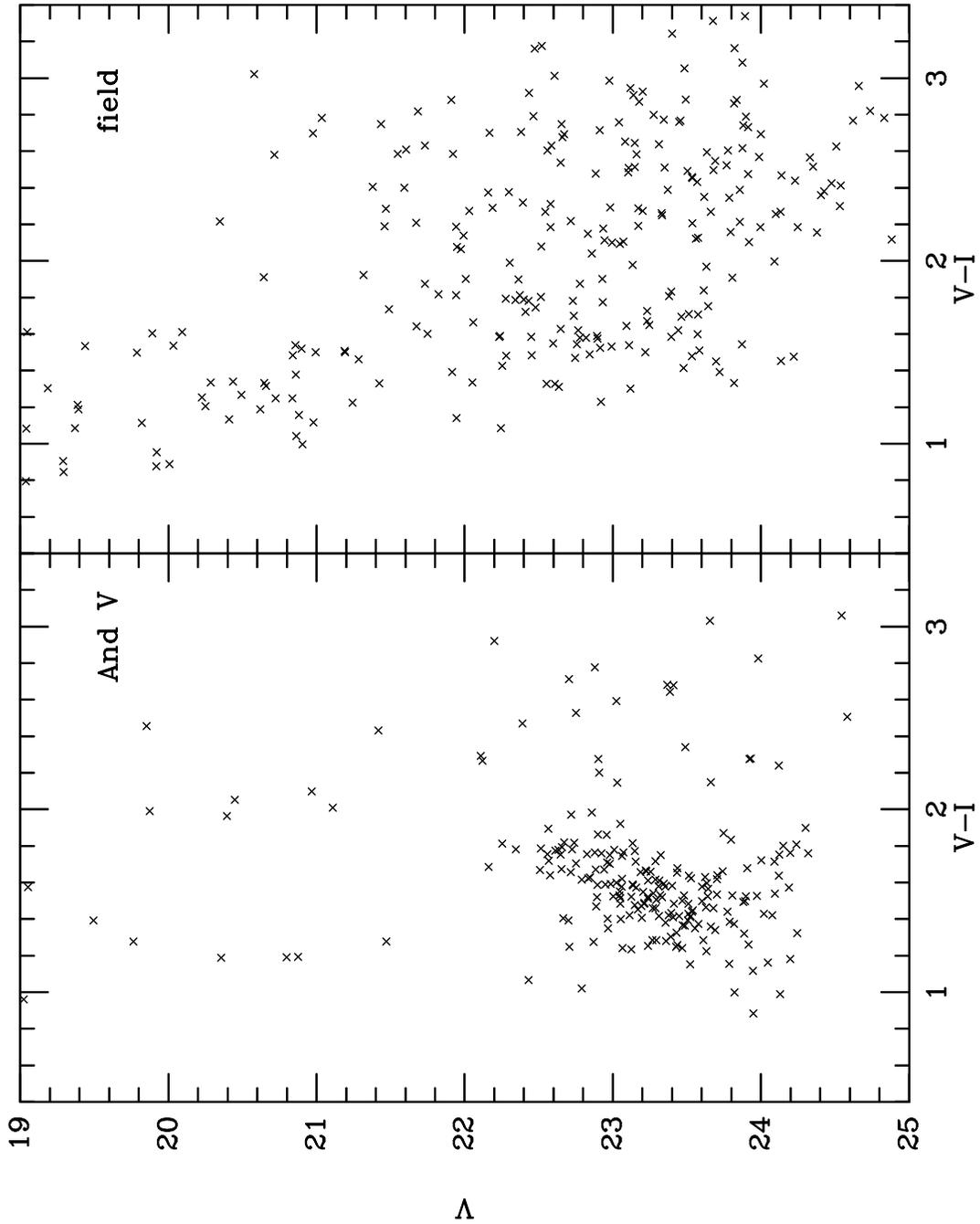}
\caption[]{Color--magnitude diagrams for stars on the And V frames.
The left panel shows stars within a radius of 170 pixels (71 arcsec) of
the center of And V, where And V members outnumber field
contamination.  The right panel displays stars greater than 400 pixels
(168 arcsec) from the center of And V, where the contribution from And
V stars is negligible.  The right panel (field) represents 5.5 times
more area on the CCD frame than the left panel (And V).  Note the upper
red giant branch that is present in the And V color--magnitude diagram,
which is absent in the field diagram.}
\label{2panelcmd}
\end{figure}

\begin{figure}[p]
\epsscale{0.70}
\plotone{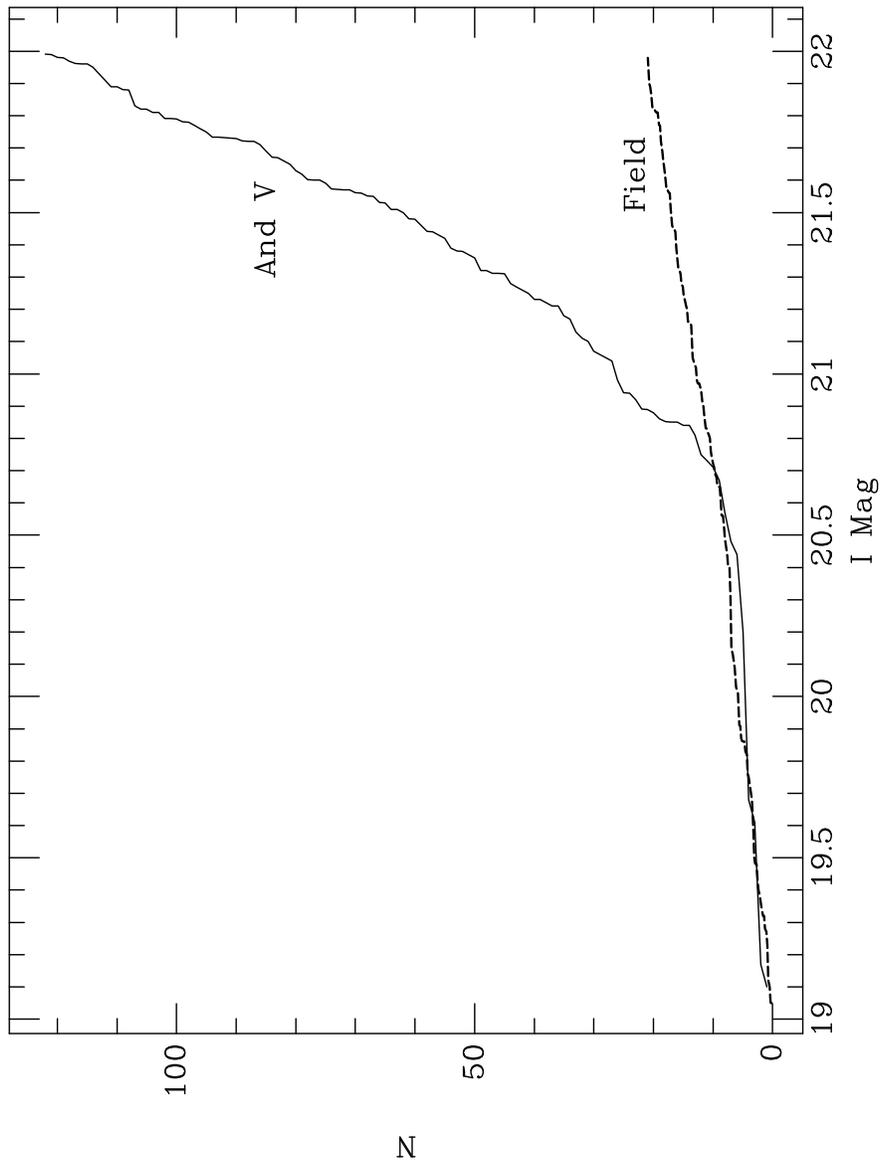}
\caption[]{Cumulative luminosity functions in $I$ for stars on the And
V frames.  The solid line shows stars within a radius of 170 pixels (71
arcsec) of the center of And V, where And V members outnumber field
contamination.  The dashed line represents stars greater than 400
pixels (168 arcsec) from the center of And V, where the contribution
from And V stars is negligible.  The field luminosity function has been
divided by 5.5 because it represents that much larger an area on the
CCD frames.  Note how the And V luminosity function strongly turns up
from the field luminosity function at $I$ = 20.85, which we interpret
as the tip of the And V red giant branch.  The use of the cumulative
luminosity function (as opposed to the differential luminosity
function) avoids binning effects and averages out small random
photometric errors.}
\label{and5_lf}
\end{figure}
 
\begin{figure}[p]
\plotone{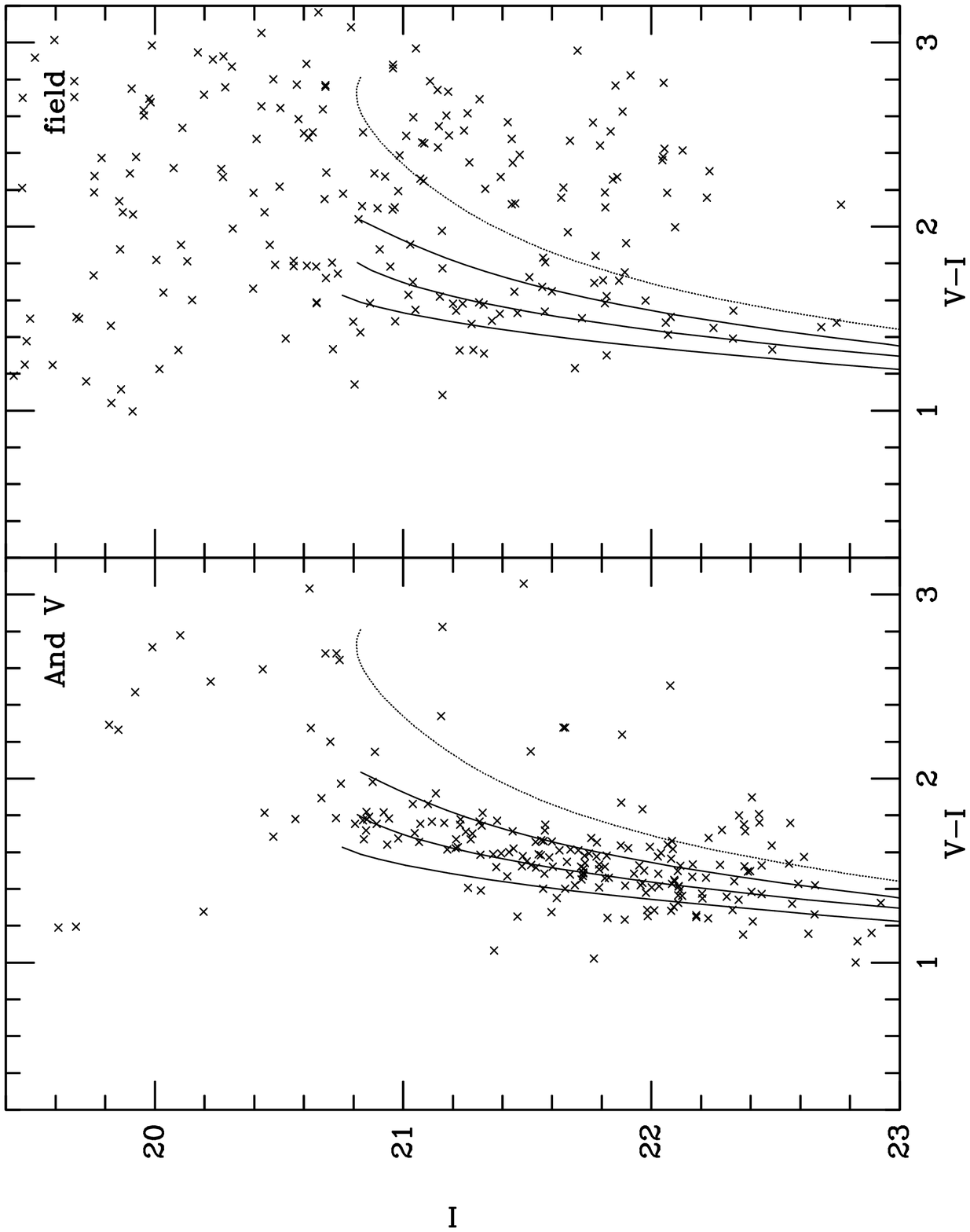}
\caption[]{Color--magnitude diagrams as in Figure \ref{2panelcmd},
except now with $I$ as the abscissa.  Red giant branch fiducials for
four Galactic globular clusters that span a range of metal abundance
(Da Costa \& Armandroff 1990), shifted to the distance modulus and
reddening of And V, have been overplotted.  From left to right, the red
giant branch fiducials are M15 ([Fe/H] = --2.17), M2 (--1.58), NGC 1851
(--1.16), and 47 Tuc (--0.71, dotted line).  The specifications of the
``And V'' and ``field'' regions are given in the text.}
\label{cmd_fiducial}
\end{figure}

\begin{references}
\reference{} Aaronson, M., Gordon, G., Mould, J., Olszewski, E., \&
Suntzeff, N.\ 1985, \apj, 296, L7
\reference{} Armandroff, T., Boroson, T., De Veny, J., Heathcote, S.,
Jacoby, G., Lauer, T., Massey, P., Reed, R., Valdes, F., \& Vaughnn,
D.\ 1998, NOAO CCD Mosaic Imager User Manual (Tucson: NOAO)
\reference{} Armandroff, T.\ E.\ 1994, in ESO/OHP Workshop on Dwarf
Galaxies, ed.\ G.\ Meylan \& P.\ Prugniel (Garching: ESO), 211
\reference{} Armandroff, T.\ E., Da Costa, G.\ S., Caldwell, N., \&
Seitzer, P.\ 1993, \aj, 106, 986
\reference{} Bertelli, G., Bressan, A., Chiosi, C., Fagotto, F., \&
Nasi, E.\ 1994, \aaps, 106, 275
\reference{} Bremmes, T., Binggeli, B., \& Prugniel, P.\ 1998, \aaps,
129, 313
\reference{} Burstein, D., \& Heiles, C.\ 1982, \aj, 87, 1165
\reference{} Caldwell, N., Armandroff, T.\ E., Seitzer, P., \& Da
Costa, G.\ S.\ 1992, \aj, 103, 840
\reference{} Condon, J.\ J., Cotton, W.\ D., Greisen, E.\ W., Yin,
Q.\ F., Perley, R.\ A., Taylor, G.\ B., \& Broderick, J.\ J.\ 1998,
\aj, 115, 1693
\reference{} Da Costa, G.\ S.\ 1998, in Stellar Astrophysics for the
Local Group: A First Step to the Universe, ed.\ A. Aparicio \&
A.\ Herrero (Cambridge: Cambridge University Press), in press
\reference{} Da Costa, G.\ S., \& Armandroff, T.\ E.\ 1990, \aj, 100,
162
\reference{} Da Costa, G.\ S., \& Armandroff, T.\ E.\ 1995, \aj, 109,
2533
\reference{} Da Costa, G.\ S., Armandroff, T.\ E., Caldwell, N., \&
Seitzer, P.\ 1996, \aj, 112, 2576
\reference{} Davies, J.\ I., Disney, M.\ J., Phillipps, S., Boyle,
B.\ J., \& Couch, W.\ J.\ 1994, \mnras, 269, 349
\reference{} Dean, J.\ F., Warren, P.\ R., \& Cousins,
A.\ W.\ J.\ 1978, \mnras, 183, 569
\reference{} Ellis, G.\ L., Grayson, E.\ T., \& Bond, H.\ E.\ 1984,
\pasp, 96, 283
\reference{} Ferguson, H.\ C., \& Binggeli, B.\ 1994, \aapr, 6, 67
\reference{} Freedman, W.\ L., \& Madore, B.\ F.\ 1990, \apj, 365, 186
\reference{} Fusi Pecci, F., Buonanno, R., Cacciari, C., Corsi, C.\ E.,
Djorgovski, S.\ G., Federici, L., Ferraro, F.\ R., Parmeggiani, G., \&
Rich, R.\ M.\ 1996, \aj, 112, 1461
\reference{} Gallagher, J.\ S., \& Wyse, R.\ F.\ G.\ 1994, \pasp, 106,
1225
\reference{} Jones, J.\ H.\ 1993, \aj, 105, 933
\reference{} Karachentsev, I.\ 1996, \aap, 305, 33
\reference{} Karachentseva, V.\ E., \& Karachentsev, I.\ D.\ 1998,
\aaps, 127, 409
\reference{} Kennicutt, R.\ C.\ 1994, in The Local Group: Comparative
and Global Properties, ed.\ A.\ Layden, R.\ C.\ Smith \& J.\ Storm
(Garching: ESO), 28
\reference{} Landolt, A.\ U.\ 1992, \aj, 104, 340
\reference{} Lasker, B.\ M., \& Postman, M.\ 1993, in ASP
Conf.\ Ser.\ 43, Sky Surveys:  Protostars to Protogalaxies,
ed.\ B.\ T.\ Soifer (San Francisco: ASP), 131
\reference{} Lee, M.\ G.\ 1995, \aj, 110, 1129
\reference{} Lee, M.\ G., Freedman, W.\ L., \& Madore, B.\ F.\ 1993,
\apj, 417, 553
\reference{} Massey, P., \& Armandroff, T.\ E.\ 1995, \aj, 109, 2470
\reference{} Massey, P., Armandroff, T., De Veny, J., Claver, C.,
Harmer, C., Jacoby, G., Schoening, B., \& Silva, D.\ 1997, Direct
Imaging Manual for Kitt Peak (Tucson: NOAO)
\reference{} Mould, J.\ 1997, \pasp, 109, 125
\reference{} Mould, J., \& Kristian, J.\ 1986, \apj, 305, 591
\reference{} Mould, J., \& Kristian, J.\ 1990, \apj, 354, 438
\reference{} Nilson, P.\ 1973, Uppsala General Catalogue of Galaxies,
Uppsala Astron.\ Obs.\ Ann.\ 6
\reference{} Pritchet, C.\ J., \& van den Bergh, S.\ 1988, \apj, 331, 135
\reference{} Reid, I.\ N., Brewer, C., Brucato, R.\ J., McKinley,
W.\ R., Maury, A., Mendenhall, D., Mould, J.\ R., Mueller, J.,
Neugebauer, G., Phinney, J., Sargent, W.\ L. W., Schombert, J., \&
Thicksten, R.\ 1991, \pasp, 103, 661
\reference{} Reid, N., \& Djorgovski, S.\ 1993, in ASP Conf.\ Ser.\ 43,
Sky Surveys:  Protostars to Protogalaxies, ed.\ B.\ T.\ Soifer (San
Francisco: ASP), 125
\reference{} Renzini, A., \& Buzzoni, A.\ 1986, in Spectral Evolution
of Galaxies, ed.\ C.\ Chiosi \& A.\ Renzini (Dordrecht: Reidel), 195
\reference{} Rice, W., Lonsdale, C.\ J., Soifer, B.\ T., Neugebauer,
G., Kopan, E.\ L., Lloyd, L.\ A., de Jong, T., \& Habing, H.\ J.\ 1988,
\apjs, 68, 91
\reference{} Sakai, S., Madore, B.\ F., \& Freedman, W.\ L.\ 1997,
\apj, 480, 589
\reference{} Schmidt, K.-H., \& Boller, T.\ 1992, Astron.\ Nachr., 313, 189
\reference{} Schombert, J.\ M., Bothun, G.\ D., Schneider, S.\ E., \&
McGaugh, S.\ 1992, \aj, 103, 1107
\reference{} Stetson, P.\ B.\ 1987, \pasp, 99, 191
\reference{} Stetson, P.\ B., Davis, L.\ E., \& Crabtree, D.\ R.\ 1990,
in ASP Conf.\ Ser.\ 8, CCDs in Astronomy, ed.\ G.\ H.\ Jacoby (San
Francisco, ASP), 289
\reference{} Trentham, N.\ 1998, \mnras, 294, 193
\reference{} van den Bergh, S.\ 1972a, \apj, 171, L31
\reference{} van den Bergh, S.\ 1972b, \apj, 178, L99
\reference{} van den Bergh, S.\ 1974, \apj, 191, 271
\reference{} van den Bergh, S.\ 1994, \apj, 428, 617
\reference{} Weinberger, R.\ 1995, \pasp, 107, 58
\reference{} Yang, H., \& Skillman, E.\ D.\ 1993, \aj, 106, 1448
\reference{} Young, L.\ M., \& Lo, K.\ Y.\ 1997, \apj, 490, 710
\end{references}
\end{document}